\documentclass[preprint, showpacs,preprintnumbers,amsmath,amssymb,pra,superscriptaddress]{revtex4}

\usepackage{graphicx}
\usepackage{dcolumn}
\usepackage{bm}

 \newcommand{\cat}[1]{|{#1}\rangle}

\begin{document}

\preprint{KOR-UNIV/QI Lab Changho}

\title{Comment on "Quantum Dialogue protocol"}

\author{Alice}
\affiliation{Department of Physics, Korea University, Seoul 133-791, Korea}
\email{alice@korea.ac.kr}
\author{Bob}
\affiliation{Center for Information Security Technologies (CIST), Korea University, Seoul 133-791, Korea}
\affiliation{Graduate School of Information Security, Korea University, Seoul 133-791, Korea}
\email{Bob@korea.ac.kr}
\author{Charlie}
\affiliation{Department of Display Semiconductor, Korea University, Yeongi-kun Choongnam 339-700, Korea}
\affiliation{Graduate School of Information Security, Korea University, Seoul 133-791, Korea}

\date{\today}

\begin{abstract}
In 2004, Ba An Nguyen [Phys. Lett. A 328, 6-10] has presented a
Quantum Dialogue scheme for simultaneously communicating their
messages. In this comment, we show that the quantum dialogue
scheme is not secure against the intercept-and-resend attack and
we propose a modified scheme which is secure against that attack.
\end{abstract}

\pacs{03.67.Dd,03.67.Hk,03.65.Ud}
\keywords{Quantum Dialogue, the intercept and resend attack}

\maketitle

In 2004, Ba An Nguyen has presented the Quantum Dialogue
scheme\cite{QD}. He demonstrated that the protocol is
asymptotically secure against the disturbance attack, the
intercept-and-resend attack and the entangle-and-measure attack.

Let us start with the brief description of the quantum dialogue
protocol. Bob initially prepares the state
$\cat{\psi_{0.0}}_{h_{n}, t_{n}}=\frac{1}{\sqrt{2}}
\left[\cat{01}+\cat{10}\right]$ and encodes his bits
$(k_{n},l_{n})$ by performing $U_{k_{n}, l_{n}} (U_{00}=I,
U_{01}=\sigma_{x},
 U_{10}=i\sigma_{y}$  and $U_{11}=\sigma_{z})$ on the state $\cat{\psi_{0.0}}_{h_{n},
 t_{n}}$. Bob keeps qubit $h_{n}$ with him and sends qubit $t_{n}$ to Alice.
Alice confirms Bob that she received qubit. She determines the
mode(the message mode(MM) or the checking mode(CM)) and encodes
her messages by performing the unitary operation $U_{i_{n},
j_{n}}$. If the mode is the checking mode, she encodes random
bits. And she sends Bob the encoded qubit $t_{n}$. Bob performs a
Bell measurement on the pair of qubits with the result in state
$\cat{\psi_{x_{n},y_{n}}}_{h_{n}, t_{n}}$, and listens to Alice to
tell him that was a run in MM or in CM. If it was a MM, Bob
publicly reveals the values of $(x_{n}, y_{n})$. Alice and Bob
decodes the each other's secret messages. That is, Alice's bits as
$i_{n}=\vert x_{n}-k_{n} \vert$ and $j_{n}=\vert y_{n}-l_{n}
\vert$ and Bob's bits as $k_{n}=\vert x_{n}-i_{n} \vert$ and
$l_{n}=\vert y_{n}-j_{n}\vert$. If it was a CM mode, Alice
publicly announces the value of $(i_{n}-j_{n})$. And Bob checks
the eavesdropping by checking both $i_{n}=\vert x_{n}-k_{n} \vert$
and $j_{n}=\vert y_{n}-l_{n}\vert$. If the checking computation is
correct, we determine that there is no eavesdropping. Otherwise
the process is discontinued.

The author claim that this protocol is asymptotically secure
against the disturbance attack, the intercept-and-resend attack
and the entangle-and-measure attack. But our simple strategy of
attack shows that the protocol is not secure against the
intercept-and-resend attack. So that, undetectable eavesdropping
scheme may exist. The method of attack is as follows.
\begin{enumerate}
\item Bob prepares initial states and encodes his messages on
the initial states. We suppose that Bob's encoded states is
$\cat{\psi_{k,l}}$. Here $(k,l)$ means a Bob's secret messages. He
stores the first photon(home photon) $h$, and sends the second
photon(travel photon) $t$ to Alice.
\item Eve intercepts the travel photon $t$ and restores it. She
generates any Bell states $\cat{\psi_{k',l'}}$ and sends the
second photon $t'$ of the states to Alice.
\item After receiving the travel photon $t'$ Alice randomly switches
between MM and CM. In the MM, Alice encodes her messages by
performing the Pauli-operation on that photon. In the CM, Alice
randomly encodes on that photon by using the same way. Then Alice
sends the photon to Bob.
\item Eve takes snatch the photon $t'$ and performs a Bell measurement on
the pair of the received photon $t'$ and the his first photon
$h'$. Then he knows the Alice's encoding operations because she
already knows Bob's $t'$ and initial states of the pairs of photon
$h'$. So, Eve successfully eavesdrops the Alice's secret messages.
And Eve performs the same operation with Alice did on the photon
$t$. Eve sends the encoded photon $t$ to Bob.
\item Bob receives the travel photon $t$ and performs a Bell
measurement on both photon $h$ and $t$ to decode Alice's
information. And Bob wait for Alice to tell him that was a perform
in MM or in CM. If it was a MM, Bob decodes the Alice's messages
by using the initial states, his encoding operation and final Bell
measurement outcome. And he publicly announce the his measurement
outcome to allow Alice also to decode Bob's message. If it was a
CM, Alice publicly announces the her operator for Bob to check the
Eve. \\
In this checking step, Eve's intervention is not reveled. Because
Eve knows Alice's messages and performs the encoding operations
according the messages.
\end{enumerate}
\begin{figure}
\includegraphics[width=12.5cm]{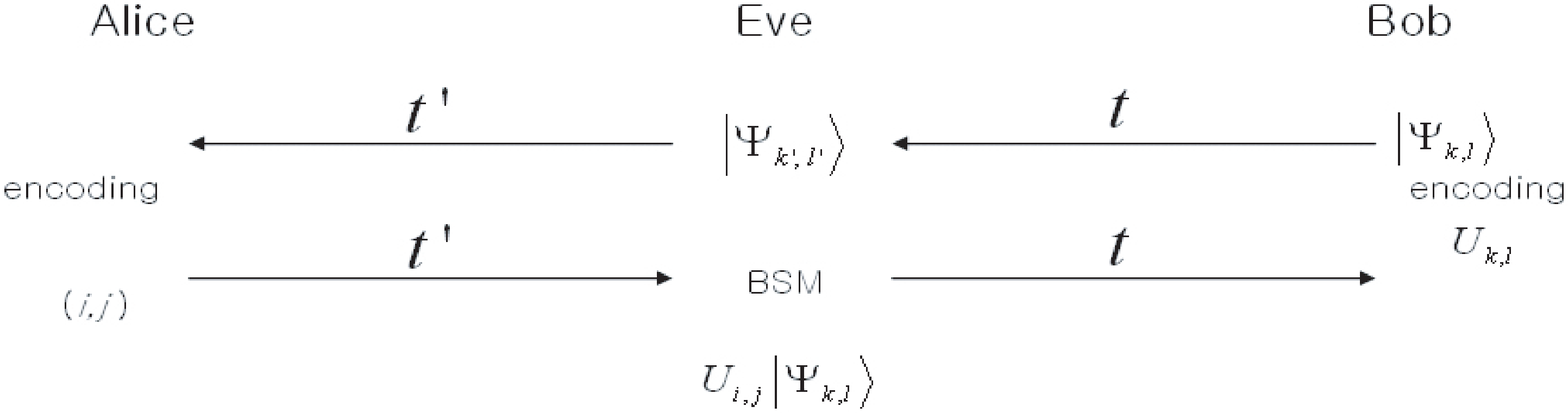}
\caption{\label{fig_1} The intercept-and-resend attack to the
quantum dialogue protocol. Here BSM means a Bell states
measurement.}
\end{figure}

Eve has no access to Bob's home photon $h$ but can handle the
travel photon $t$ while it goes from Bob to Alice and travel
photon $t'$ when from Alice to Bob. Our attacking strategy is not
detected by permitted users. Eve eavesdrops the Alice's all secret
messages without detection. But he doesn't know Bob's messages.
Note that the attacker is not revealed to the right users and the
messages of the one of users is exposed to the attacker in the
attack.

Let us now consider how to modify the Quantum Dialogue protocol to
make it secure against the proposed attack. Bob initially prepares
the state $\cat{\psi_{0.0}}_{h_{n}, t_{n}}=\frac{1}{\sqrt{2}}
\left[\cat{01}+\cat{10}\right]$ and chooses the mode of two mode,
MM and CM. For the MM, he encodes his secret messages
$(k_{n},l_{n})$ by performing $U_{k_{n}, l_{n}}$ on the state
$\cat{\psi_{0.0}}_{h_{n},t_{n}}$. For the CM, he encodes just
random bits. Then he sends the second photon $t_{n}$ to Alice.
Alice's choice of the mode is the same with the Bob's action.
Alice sends the photon $t_{n}$ back to Bob. Bob performs a Bell
measurement on the pairs of the photon $h_{n}$ and $t_{n}$. Alice
and Bob publicly announce the mode which they chose. If Alice's
and Bob's choice was CM at the same time, they inform their
encoding operation and Bob announce his measurement outcome. Then
they can determine the Eve's intervention by checking the
correlation of states. If Alice's and Bob's mode choice was
different, they do not inform their encoding operations. In this
case, only one user can know opposite user's messages. That is, If
Alice chose CM and Bob chose MM, Alice can know Bob's message by
using her operation and measurement outcome. If Alice chose MM and
Bob chose CM, Bob can know Alice's message by using Alice's
measurement outcome and his operation. If Alice's and Bob's mode
was MM at the same time, the decoding method is the same as the
original quantum dialogue protocol. The point of security of this
protocol is that each of two communicators have a choice of the
mode. The checking mode is runs if and only if they chose CM
simultaneously. If they selected different mode each other, the
protocol runs in one-way communication.

In summary, we show that the quantum dialogue protocol proposed by
Nguyen\cite{QD} is not secure against the intercept-and-resend
attack and propose a modified quantum dialogue protocol which is
secure against the attack described above. The modified quantum
dialogue protocol is asymptotically secure against the disturbance
attack, the entangle-and-measure attack and the
intercept-and-resend attack. Its proofs about asymptotic security
is the same as that of the original quantum dialogue protocol.
%
%
%
%
\bibliographystyle{siam}

\end{document}